\documentclass{pnastwo}
\usepackage[pdftex]{graphicx}
\usepackage{amssymb,amsfonts,amsmath}
\newcommand \bea {\begin{eqnarray}}
\newcommand \eea {\end{eqnarray}}
\contributor{Submitted to Proceedings
of the National Academy of Sciences of the United States of America}
\url{www.pnas.org/cgi/doi/10.1073/pnas.0709640104}
\copyrightyear{2008}
\issuedate{Issue Date}
\volume{Volume}
\issuenumber{Issue Number}

\begin{document}



\title{Competition between collective and individual dynamics}
\author{S\'ebastian Grauwin\affil{1}{Universit\'e de Lyon, Laboratoire de Physique,ENS Lyon and CNRS, 46 all\'ee d'Italie, F-69007 Lyon, France
}
\affil{2}{IXXI, 5 rue du Vercors, F-69007 Lyon, France}, 
Eric Bertin\affil{1}{}\affil{2}{},
R\'emi Lemoy\affil{2}{}\affil{3}{Laboratoire d'Economie des Transports,Universit\'e Lyon 2 and CNRS, 14, av. Berthelot F-69007 Lyon, France},
\and Pablo Jensen\affil{1}{}\affil{2}{}\affil{3}{}\affil{4}{To whom correspondence should be addressed; E-mail: pablo.jensen@ens-lyon.fr}}

\contributor{Submitted to Proceedings of the National Academy of Sciences
of the United States of America}

\maketitle

\begin{article}

\begin{abstract}
Linking the microscopic and macroscopic behavior is at the heart of many natural and social sciences. This apparent similarity conceals essential differences across disciplines: while physical particles are assumed to optimize the global energy, economic agents maximize their own utility. Here, we solve exactly a Schelling-like segregation model, which interpolates continuously between cooperative and individual dynamics. We show that increasing the degree of cooperativity induces a qualitative transition from a segregated phase of low utility towards a mixed phase of high utility. By introducing a simple function which links the individual and global levels, we pave the way to a rigorous approach of a wide class of systems, where dynamics is governed by individual strategies.
\end{abstract}

\keywords{socioeconomy | statistical physics | segregation | phase transitions}

The intricate relations between the individual and collective levels are at the heart of many natural and social sciences. Different disciplines wonder how atoms combine to form solids \cite{cambridge,goodstein}, neurons give rise to consciousness \cite{damasio,changeux} or individuals shape markets and societies \cite{econophysics,smith,latour07}. However, scientific fields assume distinct points of view for defining the "normal", or "equilibrium" aggregated state. Physics looks at the collective level, selecting the configurations that minimize the global free energy \cite{goodstein}. In contrast, economic agents behave in a selfish way, and equilibrium is attained when no agent can increase its own satisfaction \cite{mascolell}. Although similar at first sight, the two approaches lead to radically different outcomes, and the selfish strategy may prove dramatically inefficient in presence of interactions between agents. We illustrate this effect on an exactly solvable model, similar to Schelling's segregation model \cite{schelling71}. Considering individual agents which prefer a mixed environment, we study the segregated or mixed patterns that emerge at the global level. A "tax" parameter monitors continuously the agents' degree of altruism or cooperativity, i.e., their consideration of the global welfare. At low degrees of cooperativity, segregation occurs and the agents' utilities remain low, in spite of continuous efforts to maximize their satisfaction. As the altruism parameter is increased, a phase transition occurs, driving the system towards a mixed phase of maximal utility. Our approach generalizes the free energy function used in physics, allowing to predict analytically the stationary states, which required so far numerical simulations \cite{fossett-clark08}.

\section*{Model}
Our model represents in a schematic way the dynamics of residential moves in a city. For simplicity, we include one type of agent, but our results can readily be generalized to deal with agents of two "colors", as in the original Schelling model \cite{schelling71} (see below). The city is divided into $Q$ blocks ($Q >> 1$), each block containing $H$ cells or flats (Fig. \ref{config}). We assume that each cell can contain at most one agent, so that the number $n_q$ of agents in a given block $q$ ($q=1,\ldots,Q$) satisfies $n_q \le H$, and we introduce the density of agents $\rho_q=n_q/H$. Each agent has the same utility function $u(\rho_q)$, which describes the degree of satisfaction concerning the density of the block it is living in. The collective utility is defined as the total utility of all the agents in the city: $U(x)=H\sum_q \rho_q u(\rho_q)$, where $x \equiv \{\rho_q\}$ corresponds to the coarse-grained configuration of the city, i.e. the knowledge of the density of each block. For a given $x$, there is a large number of ways to arrange the agents in the different cells. This number of arrangements is quantified by its logarithm $S(x)$, called the entropy of the configuration $x$.

The dynamical rule allowing the agents to move from one block to another is the following. At each time step, one picks up at random an agent and a vacant cell, within two different blocks. Then the agent moves in that empty cell with probability:  
\be
P_{xy} = \frac{1}{1+e^{-\mathcal{C}/T}},
\ee
where $x$ and $y$ are respectively the configurations before and after the move, and $\mathcal{C}$ is the cost associated to the proposed move. The positive parameter $T$ is a "temperature" which introduces in a standard way \cite{anderson92} some noise on the decision process. It can be interpreted as the effect of features that are not explicitely included in the utility function but still affect the moving decision (urban facilities, friends\ldots).
We write the cost $\mathcal{C}$ as :
\bea \label{def-C}
\mathcal{C} &=& \Delta u + \alpha (\Delta U - \Delta u)
\eea
where $\Delta u$ is the variation of the agent's own utility upon moving and $\Delta U$ is the variation of the total utility of all agents. The parameter $0\leq \alpha\leq 1$ weights the contribution of the other agents' utility variation in the calculation of the cost $\mathcal{C}$, and it can thus be interpreted as a degree of cooperativity (or altruism). 
For $\alpha =0$, the probability to move only depends on the selfish interest of the chosen agent, while for $\alpha=1$ it only depends on the collective utility. Varying $\alpha$ in a continuous way, one can interpolate between these two limiting behaviors.

\section*{Potential function}
We wish to find the stationary probability distribution $\Pi(x)$ of the microscopic configurations $x$. If the cost $\mathcal{C}$ can be written as $\mathcal{C}=\Delta V \equiv V(y)-V(x)$, where $V(x)$ is a function of the configuration $x$, then the dynamics satisfies detailed balance \cite{evans} and the distribution $\Pi(x)$ is given by
\be
\Pi(x) = \frac{1}{Z}\, e^{F(x)/T},
\ee
with $F(x)=V(x)+TS(x)$ and $Z$ a normalization constant.
The entropy has for large $H$ the standard expression $S(x)=H\sum_q s(\rho_q)$, with
\be
s(\rho) = -\rho\ln\rho -(1-\rho)\ln(1-\rho).
\ee
We now need to find the function $V(x)$, if it exists. Given the form (\ref{def-C}) of $\mathcal{C}$, finding such a function $V(x)$ amounts to finding a "linking" function $L(x)$, connecting the individual and collective levels, such that $\Delta u = \Delta L$. By analogy to the entropy, we assume that $L(x)$ can be written as a sum over the blocks, namely $L(x)=H\sum_q \ell(\rho_q)$. Considering a move from a block at density $\rho_1$ to a block at density $\rho_2$, $\Delta L$ reduces in the large $H$ limit to $\ell'(\rho_2)-\ell'(\rho_1)$, where $\ell'$ is the derivative of $\ell$. The condition $\Delta u = \Delta L$ then leads to the identification $\ell'(\rho)=u(\rho)$, from which the expression of $\ell(\rho)$ follows:
\be
\ell(\rho) = \int_0^{\rho} u(\rho') d\rho'.
\label{link}
\ee
As a result, the function $F(x)$ can be expressed in the large $H$ limit as $F(x)=H\sum_q f(\rho_q)$, with a block potential $f(\rho)$ given by :

\bea \nonumber
f(\rho) &=& -T\rho\ln\rho - T(1-\rho)\ln(1-\rho)\\
&+& \alpha\rho u(\rho) + (1-\alpha)\int_{0}^{\rho} u(\rho')d\rho'.
\label{f-rho}
\eea

The probability $\Pi(x)$ is dominated by the configurations $x=\{\rho_q\}$ that maximize the sum $\sum_q f(\rho_q)$ under the constraint of a fixed $\rho_0 = 1/Q \sum_{q=1}^Q \rho_q$. To perform this maximization procedure, we follow standard physics methods used in the study of phase transitions (like liquid-vapor coexistence \cite{callen85}), which can be summarized as follows. If $f(\rho)$ coincides with its concave hull at a given density $\rho_0$, then the state of the city is homogeneous, and all blocks have a density $\rho_0$. Otherwise, a phase separation occurs: some blocks have a density $\rho_1^*<\rho_0$, while the others have a density $\rho_2^*>\rho_0$.

Interestingly, the potential $F$ can be calculated for arbitrary utility functions. It generalizes the free energy function used in physics, allowing to predict analytically the global town state. Such an achievement eluded so far individualistic, Schelling-type models, which had to be solved through numerical simulations \cite{fossett-clark08}.

\begin{figure}
\begin{center}
\includegraphics[width=5cm]{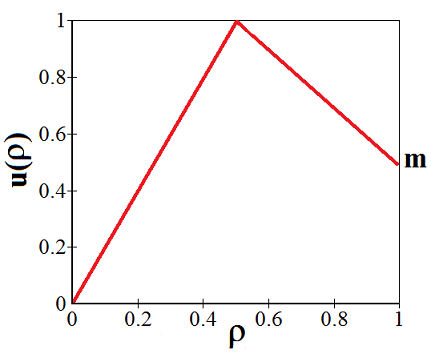}
\end{center}
\caption{Asymmetrically peaked individual utility as a function of block density. The utility is defined as $u(\rho)=2\rho$ if $\rho \leq 1/2$ and $u(\rho)=m+2(1-m)(1-\rho)$ if $\rho > 1/2$, where $0<m<1$ is the asymmetry parameter. Agents strictly prefer half-filled neighborhoods ($\rho=1/2$). They also prefer overcrowded ($\rho=1$) neighborhoods to empty ones ($\rho=0$).}
\label{um}
\end{figure}

To obtain explicitly the equilibrium configurations, one needs to know the specific form of the utility function. To illustrate the dramatic influence of the cooperativity parameter $\alpha$, we use the asymmetrically peaked utility function \cite{pancs07}, which indicates that agents prefer mixed blocks (Figure \ref{um}). The overall town density is fixed at $\rho_0 = 1/2$ to avoid the trivial utility frustration resulting from the impossibility to attain the optimal equilibrium ($\rho_q = 1/2$ for all blocks). The qualitative behavior of the system is unchanged for $\rho_0 \neq 1/2$ or for low values of the temperature.

\begin{figure}
\begin{center}
\includegraphics[width=9cm]{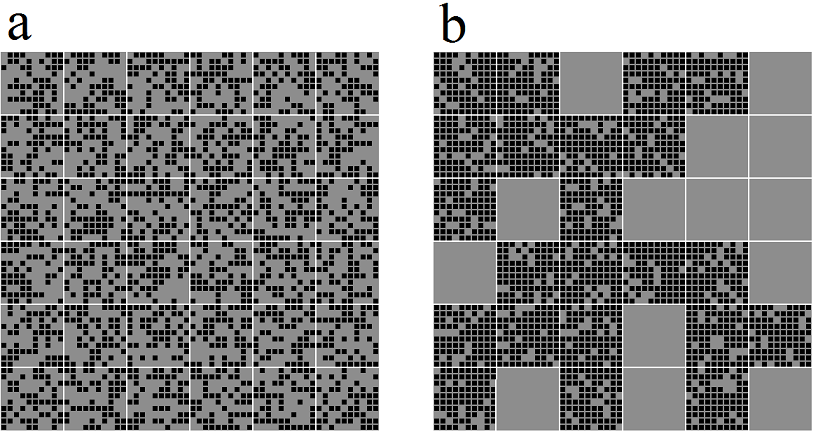}
\end{center}
\caption{Configurations of a city composed of $Q=36$ blocks containing each $H=100$ cells, with $\rho_0=1/2$. \textbf{(a) Mixed} state. Stationary state of the city for $m=0.5$, $\alpha=1$ and $T \to 0$. Agents are distributed homogeneously between the blocks, each of them having a density of $0.5$. \textbf{(b) Segregated} configuration. Stationary state of the city for $m=0.5$, $\alpha=0$ and $T \to 0$. Agents are gathered on $20$ blocks of mean density $0.9$, the other blocks being empty. In the original Schelling model \cite{schelling71}, each agent has a distinct neighborhood, defined by its 8 nearest neighbors. Here, we only keep the essential ingredient of blocks of distinct densities. Our model shows the same qualitative behavior as Schelling's but can be solved exactly, thanks to the {\it partial} reduction of agent's heterogeneity.}
\label{config}
\end{figure}

In the collective case ($\alpha=1$), the optimal state corresponds to the configuration that maximizes the global utility, which can be immediately guessed from Figure \ref{um}, namely $\rho_q = 1/2$ for all $q$ (Fig \ref{config}a). On the contrary, in the selfish case ($\alpha=0$, Fig \ref{config}b), maximization of the potential $F(x)$ shows that the town settles in a segregated configuration where a fraction of the blocks are empty and the others have a density $\rho_s > 1/2$. Surprisingly, the city settles in this state of low utility in spite of agents' continuous efforts to maximize their own satisfaction. To understand this frustrated configuration, note that the collective equilibrium ($\rho_q = 1/2$ for all $q$) is now an {\it unstable} Nash equilibrium at $T > 0$. The instability can be understood by noting that at $T > 0$ there is a positive probability that an agent accepts a slight decrease of its utility, and leaves a block with density $\rho_q = 1/2$. The agents remaining in its former block now have a lower utility and are more likely to leave. This creates an avalanche which empties the block, as each move away further decreases the utility of the remaining agents.

\begin{figure}
\begin{center}
\includegraphics[width=8.5cm]{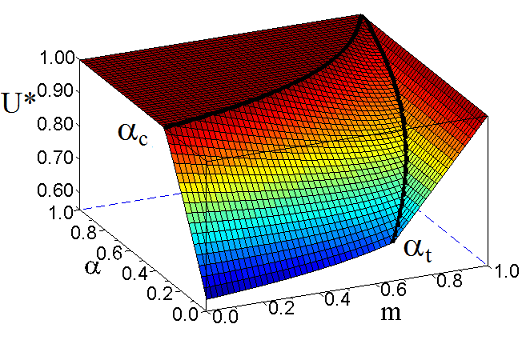}
\end{center}
\caption{Phase diagram of the global utility as a function of the cooperativity $\alpha$ and the asymmetry $m$, at $T\to 0$ and $\rho_0=1/2$. The average utility per agent $U^* = U / (H \rho_0)$ is calculated by maximizing the potential $F(x)$ for the peaked utility shown in Fig.~\ref{um}. The plateau at high values of $\alpha$ corresponds to the mixed phase of optimal utility, which is separated from the segregated state by a phase transition arising at $\alpha_c= 1/(3-2m)$. The overall picture is qualitatively unchanged for low but finite values of the temperature.}
\label{dp}
\end{figure}

Mixed and segregated states are separated by a phase transition at the critical value $\alpha_c = 1/(3-2m)$ (Figure \ref{dp}). This transition differs from standard equilibrium phase transitions known in physics, which are most often driven by the competition between energy and entropy. Here, the transition is driven by a competition between the collective and individual components of the agents' dynamics. The unsatisfactory global state of the city can be interpreted, from the economics' point of view, as an effect of externalities: by moving to increase its utility, an agent may decrease other agents' utilities, without taking this into account. From a standard interpretation in terms of Pigouvian tax \cite{pigou}, one expects that $\alpha=1$ is necessary to reach the optimal state, since by definition this value internalizes all the externalities the agent causes to the others when moving. Our results show that the optimal state is maintained until much lower tax values (for example, $\alpha_c = 1/3$ at $m=0$), a surprising result which deserves further analysis. Another interesting effect is observed for $m>2/3$ (Figure \ref{dp}). Introducing a small tax has no effect on the overall satisfaction, the utility remaining constant until a threshold level is attained at $\alpha_t= (3m-2)/(6-5m)$.\\

\begin{figure}
\begin{center}
\includegraphics[width=9.5cm]{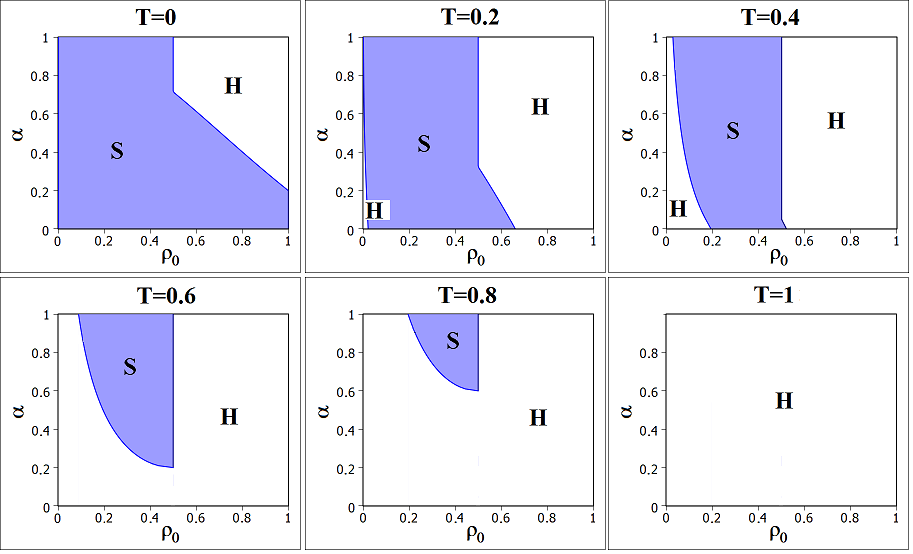}
\end{center}
\caption{Phase diagrams for the asymmetrically peaked individual utility (Fig.~\ref{um}, with $m=0.8$) for different values of $T$. Increasing the temperature $T$ tends to favor homogeneous states (white regions labelled "H") over segregated phases (grey regions labelled "S"). For small but finite temperatures (roughly $T<0.2$), the phase diagram is modified only for extremal values of $\rho_0$, as expected from the entropic term $Ts(\rho)=-T\rho \ln\rho -T (1-\rho)\ln(1-\rho)$. As $T$ is increased, the whole diagram is affected by the entropic term. Compared to the $T=0$ case, the main change is the appearance of a second homogeneous phase for $\rho_0 < 1/2$. But while for $\rho_0> 1/2$ homogeneity corresponds to the optimal choice for the agents, for $\rho_0 < 1/2$, collective utility is not maximized in a homogeneous city. The city is homogeneous by noise, not by choice.}
\label{ra-T}
\end{figure}

We focused up to now on the zero temperature limit. For low temperatures, the main qualitative conclusions are not modified, as the phase diagram is modified only for extremal values of $\rho_0$ by entropic contributions. At higher temperatures the city tends to become homogeneous, as the effect of "noise" (i.e., of the features that are not described in the model) dominates over the utility associated to density of the blocks (see Fig.~\ref{ra-T}).

\section*{Link to Schelling's original segregation model}
There are two main differences between our simple model and Schelling's original model \cite{schelling71} : the existence of agents of {\em two} colors and the definition of the agents' neighborhoods. We now show that these additional features do not introduce any essential effect.

Let us start by introducing agents of two "colors" (such as red and green). Simple calculations show that for two species which only care about the density of neighbors of their own color, the block potential (eq. \ref{f-rho}) becomes : 

\bea
f(\rho_R,\rho_G) &=& -T\rho_R\ln\rho_R -T\rho_G\ln\rho_G\nonumber\\
&-& T(1-\rho_R -\rho_G)\ln(1-\rho_R -\rho_G) \nonumber\\ 
&+& \alpha\Big[\rho_R \,u_R(\rho_R) + \rho_G \,u_G(\rho_G)\Big]\nonumber\\ 
&+& (1-\alpha)\Big[\int_{0}^{\rho_R}u_R(\rho')d\rho' + \int_{0}^{\rho_G}u_G(\rho')d\rho'\Big] \nonumber
\eea
with straightforward notations (for example $u_R(\rho_R)$ represents the utility of a red agent in a block with a density $\rho_R$ of red agents).

Finding the equilibrium configurations amounts to finding the set $\{\rho_{qR},\rho_{qG}\}$ which maximizes the potential $F(x)=\sum_q f(\rho_{qR},\rho_{qG})$ with the constraints $\sum_q\rho_{qR}=Q\rho_{0R}$ and $\sum_q\rho_{qG}=Q\rho_{0G}$, where $\rho_{0G}$ and $\rho_{0R}$ represent respectively the overall concentration of green and red agents. 

\begin{figure}
\begin{center}
\includegraphics[width=7.5cm]{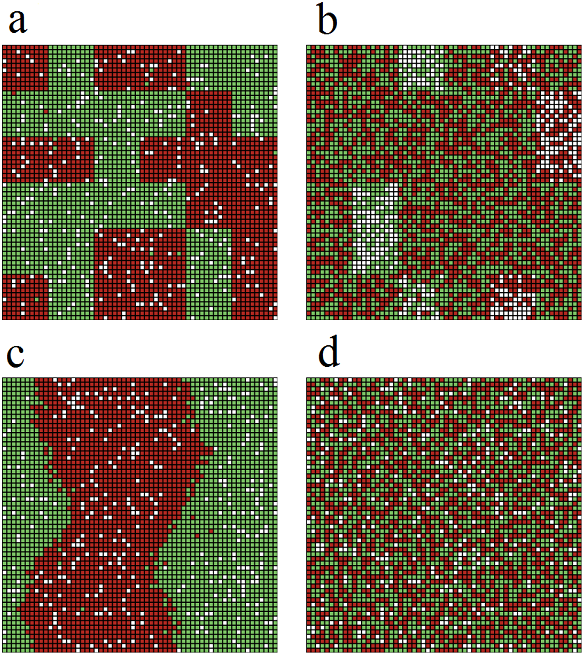}
\end{center}
\caption{The transition from segregated to mixed states is not affected by the choice of the neighborhood's definition. The figures represent stationary configurations obtained by simulating a city containing an equal number of red and green agents, whose utilities are given by the asymmetrically peaked utility function shown in Fig.~\ref{um} ($m=0.5)$. 
\textbf{Top panel.} The city is divided into blocks of size $H=100$. As predicted by the analytic model, a segregated configuration is obtained for $\alpha=0$ (a) and a mixed configuration for $\alpha=1$ (b). \textbf{Bottom panel.} The utility of an agent depends on the local density of similar neighbors computed on the $H=108$ nearest cells. The observed phases are identical to those obtained with the block description : a segregated configuration for $\alpha=0$ (c) and a homogeneous configuration for $\alpha=1$ (d). In all the simulations, we fix the proportion of vacant cells (in white) to $10\%$ and the temperature to $0.1$. This small amount of noise does not change significantly the nature of the stationary states compared to the case $T \to 0$, but reduces the simulation convergence time.}
\label{2pop}
\end{figure}

Because of the spatial constraints (the densities of red and green agents in each block $q$ must verify $\rho_{qR} + \rho_{qG} \leq 1$), the `two populations' model cannot formally be reduced to two independent `one population' models. However, the stationary states can still be computed. Let us focus once again on the $T \to 0$ limit and suppose for example that $\rho_{0R}=\rho_{0G}=\rho_0/2$. The stationary states depend once again on the values of $\rho_0$, $m$ and $\alpha$. For low values of $\alpha$, it can be shown that the system settles in a segregated state where each block contains only one kind of agent with a density $\rho_0$ (see Figure \ref{2pop}a). For $\alpha \geq \alpha_c$, the system settles in a mixed state where the density of a group in a block is either $0$ or $1/2$ (Figure \ref{2pop}b).

We now turn to the difference in agent's neighborhoods. In Schelling's original model, agents' neighbors are defined as their 8 nearest neighbors. Our model considers instead predefined blocks of common neighbors. First, it should be noted that there is no decisive argument in favor of either neighborhood definition in terms of the realism of the description of real social neighborhoods. Second, we note that introducing blocks allows for an analytical solution for arbitrary utility functions. This contrasts with the nearest neighbor case, where the best analytical approach solves only a modified model which abandons the individual point of view and is limited to a specific utility function \cite{zhang}. Finally, the simulations presented on Figures \ref{2pop} show that the transition from segregated to mixed states is not affected by the choice of the neighborhood's definition. We conclude that the block description is more adapted to this kind of simple modelling, which aims at showing stylized facts as segregation transitions.

\section*{Conclusion}
Our simple model raises a number of interesting questions about collective or individual points of view. In the purely collective case ($\alpha=1$), the stationary state corresponds to the maximization of the average utility, in analogy to the minimization of energy in physics. In the opposite case ($\alpha=0$), the stationary state strongly differs from the simple collection of individual optima \cite{represent}: the optimization strategy based on purely individual dynamics fails, illustrating the unexpected links between micromotives and macrobehavior \cite{schelling78}. However, the emergent collective state can be efficiently captured by the maximization of the linking function $\ell(\rho)$ given in Eq.~(\ref{link}), up to constraints in the overall town density. This function intimately connects the individual and global points of view. First, it depends only on the global town configuration (given by the $\rho_q$), allowing a relatively simple calculation of the equilibrium. At the same time, it can be interpreted as the sum of the {\it individual} marginal utilities gained by agents as they progressively fill the city after leaving a reservoir of zero utility. In the stationary state, a maximal value of the potential $L$ is reached. This means that no agent can increase its utility by moving (since $\Delta u = \Delta L$), consistently with the economists' definition of a Nash equilibrium. 

Equilibrium statistical mechanics has developed powerful tools to link the microscopic and macroscopic levels. These tools are limited to physical systems, where dynamics is governed by a global quantity such as the total energy. By introducing a link function, analogous to state functions in thermodynamics or potential functions in game theory \cite{monderer96}, we have extended the framework of statistical mechanics to a Schelling-like model. Such an approach paves the way to analytical treatments of a much wider class of systems, where dynamics is governed by individual strategies.

\begin{acknowledgments}
We acknowledge interesting discussions with Florence Goffette-Nagot (Groupe d'Analyse et de Th\'eorie Economique, Lyon).
\end{acknowledgments}

\end{article}


\begin{thebibliography}{10}

\bibitem{cambridge}
Cotterill, R.
\newblock (2008) {\em The Cambridge Guide to the Material World}.
\newblock (Cambridge University Press).

\bibitem{goodstein}
Goodstein, D.
\newblock (1985) {\em {States of Matter}}.
\newblock (Dover).

\bibitem{damasio}
Damasio, A.~R.
\newblock (1995) {\em The Feeling of What Happens}.
\newblock (Harcourt Brace and Company, New York).

\bibitem{changeux}
Changeux, J.~P.
\newblock (2009) {\em The Physiology of Truth: Neuroscience and Human
  Knowledge}.
\newblock (The Belknap Press).

\bibitem{econophysics}
Mantegna, R.~N. and Stanley H.~E.
\newblock (2004) {\em An introduction to econophysics : correlations and complexity in finance}.
\newblock (Cambridge Univ. Press).
\bibitem{smith}
Smith, A.
\newblock (1776) {\em {An Inquiry into the Nature and Causes of the Wealth of
  Nations}}.
\newblock (W. Strahan and T. Cadell, London).

\bibitem{latour07}
Latour, B.
\newblock (2007) {\em {Reassembling the Social: An Introduction to
  Actor-Network-Theory}}.
\newblock (Oxford University Press).

\bibitem{mascolell}
Mas-Colell, A, D., W.~M,  \& R., G.~J.
\newblock (1995) {\em Microeconomic Theory}.
\newblock (Oxford University Press, New York).

\bibitem{schelling71}
Schelling, T.~C.
\newblock (1971) {\em Journal of Mathematical Sociology} {\bf 1}, 143--186.

\bibitem{fossett-clark08}
Clark, W \& Fossett, M.
\newblock (2008) {\em Proceedings of the National Academy of Sciences} {\bf
  105}, 4109-4114.

\bibitem{anderson92}
Anderson, S.~P, De~Palma, A,  \& Thisse, J.~F.
\newblock (1992) {\em {Discrete Choice Theory of Product Differentiation}}.
\newblock (MIT Press).

\bibitem{evans}
Evans, M \& Hanney, T.
\newblock (2005) {\em J. Phys. A: Math. Gen} {\bf 38}, R195--R240.

\bibitem{callen85}
Callen, H.
\newblock (1985) {\em {Thermodynamics and an introduction to
  thermostatistics}}.
\newblock (J. Wiley and sons, New York).

\bibitem{pancs07}
Pancs, R \& Vriend, N.
\newblock (2007) {\em Journal of Public Economics} {\bf 91}, 1--24.

\bibitem{pigou}
Auerbach, A.
\newblock (1985) {\em Handbook of Public Economics} {\bf 1}, 61--127.

\bibitem{zhang}
Zhang, J.
\newblock (2004) {\em Journal of Economic Behavior and Organization}, {\bf 54}, 533--550.

\bibitem{represent}
Kirman, A.~P.
\newblock (1992) {\em Journal of Economic Perspectives} {\bf 6}, 117--36.

\bibitem{schelling78}
Schelling, T.~C.
\newblock (1978) {\em Micromotives and macrobehavior}.
\newblock (WW Norton).

\bibitem{monderer96}
Monderer, D \& Shapley, L.
\newblock (1996) {\em Games and Economic Behavior} {\bf 14}, 124--143.

\end{thebibliography}
\end{document}